\documentstyle[titlepage,12pt]{article}

\newcommand{\beq}{\begin{equation}}
\newcommand{\eeq}{\end{equation}}
\newcommand{\crossed}[1]{#1\hspace{-.5em}\backslash}
\newcommand{\eq}[1]{eq.(\ref{#1})}

\title {
 THE LAMB SHIFT OF HYDROGEN AND
LOW-ENERGY TESTS OF QED\thanks{A talk presented at  seminars
at the Max-Plank-Institut f\"ur
Quantenoptik (November, 15, 1994) and at the Physikalisches Institut
(the Universit\"at Heidelberg, November, 17, 1994).
}
}

\author {Savely G. Karshenboim \medskip\\
D. I.  Mendeleev Institute for Metrology (VNIIM),\\  St. Petersburg 198005,
Russia}

\begin{document}

\maketitle

\begin{abstract}

Leading
logarithmic corrections to the difference of Lamb shifts of s-states
$E_L(1s)-8E_L(2s)$ and to the life time of $2p_{1/2}$ state are
considered.
The result of Sokolov and Yakovlev for the Lamb splitting of $2s_{1/2}$ and
$2p_{1/2}$ is re-evaluated and our new value is 1057.8576(21) MHz. Using
 value of $E_L(1s)-8E_L(2s)=$ -187.237(8) MHz, obtained here, a
new value of the Hydrogen
and Deuterium ground states using all recent measurements connected
with 1s or 2s Lamb shifts. The highest precision value for
the Hydrogen ground state is
obtained from corrected result of Sokolov and Yakovlev experiment
 as 8172.934(22) MHz.

The Deuterium result is $E_L(1s)-8E_L(2s)=$ -187.229(8) MHz,  using it
we obtain from the Garching experiment 8183.905(224) MHz.

Some related topics are also considered. Taking into account nearest future
result the fine structure constant problem are reviewed. Special attention is
payed to Muonium hyperfine splitting, which it is done simillar to
Lamb shift calculatuion for and to the
neutron de Broglie's wave length measurements which also could lead to
some connection between the Deuterium mass and the fine structure constant.

\end{abstract}

\section{Introduction}

There is a lot of precision results of atomic Hydrogen spectrum,
which include 1s Lamb shift, 2s or both of them.
The first part of this work is devoted the problem how to obtain
a connection between these two Lamb shift values. The second part of the work
is devoted to problem of the fine structure constant determination and
some discussion on QED calculation of corrections to the Muonium
hyperfine splitting and Hydrogen Lamb shift.

The main point is theoretical calculation of value

\[
\left(
\Delta E_L(1s_{1/2}) - 8 \Delta E_L(2s_{1/2})
\right).
\]

Using this value it is possible

(i) to evaluate data included 1s and 2s Lamb shift both

(ii) to re-calculate 1s shift to 2s or 2s to 1s.

Lower as an instance we are going to consider how to obtain 1s  from 2s.
We will consider only Hydrogen, but this difference has the same expession also
for Deuterium and Muonium. Using Deuterium Lamb shift measurement,
Isotop shift of $2s \to 1s$
 and Hydrogen 2s-1s measurement it is possible to obtain Rydberg
constant without any nuclear corrections calculations or measurements.

\section{1s Lamb shift from 2s}

Recently a new result for the two-photon transition $2s \to 1s$ in the
Hydrogen has been obtained \cite{And}. To determinate the Rydberg constant
from this transition frequency it has to know the Lamb shift of the ground
level.  Neither the best direct experimental results nor theoretical one
are capable to be used without precision lowering.

The most precise value can be evaluated from the follow equation
\cite{YF1,newkar}

\[
\Delta E_L(1s_{1/2})=
\left(
\Delta E_L(1s_{1/2}) - 8 \Delta E_L(2s_{1/2})
\right)_{th}
\]

\[
+
8\left(
E(2s_{1/2})- E(2p_{1/2})
\right)_{exp}
\]

\beq                 \label{way}
+8\left(\Delta E_L(2p_{1/2})
+\Delta E_{BG}(2p_{1/2})
\right)_{th},
\eeq

where items with the indexes $th$ and $exp$ should be obtained
theoretically and experimentally, respectively, and $\Delta
E_{BG}(2p_{1/2})$ is the known correction of Barker and Glover, arisen from
the Breit equation (\cite{Bar}).

\section{Bethe logarithm and natural relativistic parameter}

As it is known, the  Bethe logarithm is

\[
\ln{k_0(n,l=0)} \approx 3.
\]

That means
that main concribution to it is due to continious spectrum states with
energies about $k_0(n,l=0)\cdot Ry$ and with momenta like

\[
\sqrt{k_0(n,l=0)} Z\alpha m\approx (4.5\div 4)Z\alpha m,
\]

where we use relativistic units:
$\hbar = c = 1$.

So the natural
relativistic parameter,
which is due to expansion of the "Dirac-Bethe logarithm" is
$\approx 4 Z\alpha$. In some meaning the natural
logarithm for Lamb shift is

\[
\ln{\frac{1+k_0(n,l=0) (Z\alpha)^2}{k_0(n,l=0) (Z\alpha)^2}}.
\]

That is the reason, why there are  large
numerical  values in a one-loop self-energy contribution.

But in this case the momenta of integration are
numericaly larger than
atomic momentum and hence a large part of contribution sould be
proportional
to square of wave function in the origin (i. e. to factor
$\delta_{l0}/n^3$).

\section{The main advantages of this calculation}

\indent

(i) Contributions of the order $\alpha^2 (Z\alpha)^5m$ to the Lamb shift
are  not obtained, but they are equal to zero
for $2p_{1/2}$ level and for s-states difference in \eq{way}.

(ii) Corrections in the order $\alpha^3 (Z\alpha)^4m$ are known
only for {\underline {$2p_{1/2}$}} state, but they are
no corrections in this order to the s-states
difference.

(iii) The precision of determination of $\alpha(Z\alpha)^6 m$ and
$\alpha(Z\alpha)^7 m$ depends on the extrapolation procedure.
 There are some large
numerical cancellation of their values in the s-states
difference and their values in the $2p_{1/2}$ state are much
smaller than s-state ones, and having the smaller values
some better approximations and higher precision can be obtained. Some
higher order corrections in higher than $\alpha(Z\alpha)^7 m$ order are
able to be canceled and extrapolation equation can include
fewer term and use fewer numerical results.

(iv) There is discrepancy between two proton charge radius
measurements (see, e. g., \cite{Han,Sim}), but the nuclear size
corrections are also equal to zero in the s-difference and
in the p-state energy.

(v) Some contributions of higher order can be important. The
leading correction is \cite{kar}

\beq   \label{cub}
\delta E^{cub}_L(ns)=-\frac{8}{27
}\frac{\alpha^2 (Z\alpha)^6m}{
\pi^2
n^3}\ln^3{\frac{1}{(Z\alpha)^2}},
\eeq

or -3.6 kHz for 2s, -29 kHz for 1s Lamb shift.

The cube logarithm term is canceled for the s-difference and
there are neither cube nor
square ones in the p-level energy expression.

\section{General expression of s-state Lamb shift difference}

See e. g. \cite{Sap}.

\beq   \label {LS}
\Delta E_L(1s_{1/2}) - 8 \Delta E_L(2s_{1/2})
=
\frac{\alpha(Z\alpha)^4}{\pi}
\frac{m_R^3}{m^2}
\eeq
\medskip

\[
\Big\{- \frac{4}{3}
\ln{\frac{k_0(1s)}{k_0(2s)}}
\left(
1+Z\frac{m}{M}
\right)^2
+(
Z\alpha)^2
\]
\medskip

\[
\left((4\ln{2}-\frac{197}{60})\ln{\frac{1}{(Z\alpha)^2}}+
(\frac{4}{15}\ln{2}+
\frac{1}{140})+G_{s}(Z\alpha)
\right)
\Big\}
\]
\medskip

\[
-\frac{7}{3}\frac{(Z\alpha)^5}{\pi}
\frac{m_R^3}{mM}
\left(\frac{3}{2}-2\ln{2}
\right)
\]
\medskip

\[
+
\frac{\alpha^2 (Z\alpha)^6m}{
\pi^2}
\left(\ln^2{\frac{1}{(Z\alpha)^2}}A_{262}
+ O(\ln{\frac{1}{(Z\alpha)^2}})
\right),
\]

\medskip

where $\ln{k_0(nl)}$ is the Bethe logarithm, $G_s$
is one-loop self-energy correction in the order $\alpha(Z\alpha)^6 m$
and higher and $A_{262}$ is the leading logarithmic two-loop correction
coefficient.

\section{General expression of the $2p_{1/2}$ state Lamb shift}

\beq   \label {LP}
\Delta E_L(2p_{1/2})=
 \frac{\alpha(Z\alpha)^4}{8\pi}
\frac{m_R^3}{m^2}
\Big\{
- \frac{4}{3}
\ln{k_0(2p)}
\left(
1+Z\frac{m}{M}
\right)^2
\eeq

\[
+(Z\alpha)^2
\left( \frac{103}{180}\ln{\frac{1}{(Z\alpha)^2}}-
\frac{9}{140}+G_{2p_{1/2}}(Z\alpha)
\right)
\Big\}
\]

\[
-\frac{7}{3}\frac{(Z\alpha)^5}{8\pi}
\frac{m_R^3}{mM}(\frac{1}{6})
+ \frac{(Z\alpha)^4}{8}
\frac{m_R^2}{m}
(
- \frac{1}{3})
\left(
\frac{1}{2}\frac{\alpha}{\pi} - 0.3285 (\frac{\alpha}{\pi})^2
+ 1.18(\frac{\alpha}{\pi})^3
\right)
\]

\[
+
O(\frac{\alpha^2 (Z\alpha)^6m}{
\pi^2}\ln{\frac{1}{(Z\alpha)^2}})
+
O(\frac{(Z\alpha)^6m^2}{M})
+,
\]

where $\ln{k_0(nl)}$ is the Bethe logarithm, $G_{2p_{1/2}}$
is one-loop self-energy corrections in the order $\alpha(Z\alpha)^6 m$
and higher.

\section{One-loop corrections}

To evaluate corrections new numerical results of one-loop self-energy
contribution to the Lamb shift \cite{Moh92} have been used.  After
subtracting all known contributions the rest can be extrapolated to $Z=1$.
We expect that corrections in the order of
$\alpha(Z\alpha)^7\ln(Z\alpha)m$ are proportional to value
$\delta_{l0}/n^3$ and our extrapolation equation for $G_{s}$ and
$G_{2p_{1/2}}$

\beq   \label {AAA}
G_{nlj}(Z\alpha)=
A_{60}(nlj)+(Z\alpha)~A_{70}(nlj).
\eeq

leads to
results \cite{YF1,newkar}

\beq         \label{Gs1}
G_s(\alpha)=0.865(21)
\eeq

and

\beq         \label{Gp1}
G_{2p_{1/2}}(\alpha)=-0.936(14).
\eeq

The value of \eq{Gs1} is in agreement with one of \cite{Pac}.

\section{Two-loop corrections}

The leading two-loop contribution to the s-state difference in \eq{way}
leads from the follow expression

\beq   \label{sig}
\delta E_L(nlj)=\langle nljm\vert
\Sigma_C(E_{nlj})\overline{G}_C(E_{nlj})
\Sigma_C(E_{nlj})\vert nljm\rangle,
\eeq

where $\Sigma_C(E)$ is one-loop self-energy
operator of an electron in the Coulomb field in the
Fried-Yennie gauge, $\vert nljm\rangle $ and $E_{nlj}$ are wave
functions and energies in the Dirac Atom of the Hydrogen, and
$\overline{G}_C(E)$ is reduced Coulomb Green function of an electron.

The
result of evaluation of \eq{sig} in
logarithmic
approximation is \cite{YF2,newkar}

\[
\delta E_L(1s_{1/2})-8\delta E_L(2s_{1/2})=\frac{\alpha^2 (Z\alpha)^6m}{
\pi^2}
\ln^2{\frac{1}{(Z\alpha)^2}}A_{262}
,
\]

where

\beq   \label{ddd}
A_{262}
=
-\frac{8}{9
}
\left(
3-2\ln{2}
\right).
\eeq

\section{Corrections to the $2p_{1/2}$-level life-time}

The most precise experimental result of the Lamb splitting
of the Hydrogen levels n=2 can be obtained from measurements \cite{Sok} of
ratio of this splitting and the radiative width of
$2p_{1/2}$ level.

The main contribution to the life time of $2p_{1/2}$ state is due to the
dipole transition and in this approximation the width of the level has form
\medskip

\beq                  \label{dip}
\Gamma_0=\frac{4 \omega^3}{3}\vert
{\bf d}_{12}\vert^2,
\eeq
\medskip

where $\omega$ is the $2p \to 1s$ transition frequency, and
${\bf d}_{12}$ is the dipole matrix element.

Relativistic corrections in order $(Z\alpha)^2$ have been obtained in
\cite{Sok}. Later in \cite{Pal} some contributions in the order
$\alpha(Z\alpha)^2$
have been also considered. The result of these works
is

\beq                  \label{pal}
\Gamma=\Gamma_0
\Big(1 +
(Z\alpha)^2
\big(
\ln{\frac{9}{8}}
-
\frac{32\alpha}{3\pi}
\big(
\ln{\frac{1}{(Z\alpha)^2}}-2.34
\big)\big)\Big).
\eeq

Considering
radiative
 corrections to \eq{dip} it should be
mentioned that there are two kinds of ones there. A
part of
corrections is due to transition frequency
shifts (i. e. Lamb shifts of
$1s_{1/2}$ and $2p_{1/2}$ levels) and the
other arises from the dipole
matrix element. One can see that the first of them leads to \eq{pal},
which really includes only a
 part of $\alpha(Z\alpha)^2$ corrections.

The other are evaluated by me \cite{newkar,YF3} in
logarithmic
approximation.
In the
Fried-Yennie gauge the terms with $\ln{Z\alpha}$ can originate
only from $1s_{1/2}$ wave function correction

\beq                  \label{wf}
\delta \psi_{1s}({\bf r})=\sum_{q\neq 1}
\psi_{qs}({\bf r})
\frac
{\langle qs \vert \Sigma_C^{(1)}
\vert 1s \rangle}
{E_{1s}-E_{qs}}
,
\eeq

but not from dipole operator or $2p_{1/2}$ wave function ones. In this
equation $\psi(r)$ is Coulomb-Schr\"odinger wave function in the coordinate
representation and the sum has to be done over all discrete and continuous
states.

The logarithmic contribution to the width is

\beq   \label{gaf}
\delta\Gamma_{wf}=\Gamma_0 \cdot
\frac{4}{3\pi}\alpha (Z\alpha)^4m \ln{\frac{1}{(Z\alpha)^2}}
\sum_{q\neq 1}
\left(
\frac{\psi_{qs}(0)}{\psi_{1s}(0)}
\right)
\frac{1}{E_{1s}-E_{qs}}
2
\frac{{\bf d}_{q2}}{{\bf d}_{12}}
,
\eeq

where the sum is over every discrete and continuous states.

After evaluation of \eq{gaf} we obtain for whole radiative correction

\beq   \label{newgam}
\delta\Gamma_{rad}=-\Gamma_0 \cdot
\frac{16}{3\pi}\alpha (Z\alpha)^2 \ln{\frac{1}{(Z\alpha)^2}}
\left(0.49158
...
 \right).
\eeq

This resulhas been also obtained analyticaly\footnote{The analytical expression
was first obtained by K. Packucki (unpublished) which use an alsolutly
different way of
calculation.} from \eq{wf}

\beq   \label{newga}
\delta\Gamma=-\Gamma_0 \cdot
\frac{16}{3\pi}\alpha (Z\alpha)^2 \ln{\frac{1}{(Z\alpha)^2}}
\left(
2-\frac{1}{2}\left(
\ln{\frac{1}{2}} + \frac{1}{16}  + \frac{8}{3}
 \right)
 \right).
\eeq

\section{Results for the atomic Hydrogen}

Let us consider items in \eq{way}. Using the results above we
find for
theoretical contributions to this equation

\beq   \label {LSn}
\Delta E_L(1s_{1/2}) - 8 \Delta E_L(2s_{1/2})
=-187.237(8)
{}~MHz
\eeq
\medskip

and
\medskip

\beq   \label {LPn}
\Delta E_L(2p_{1/2})=-12.8385(15)~MHz.
\eeq
\medskip

The two-logarithmic two-loop corrections of \eq{ddd} to \eq{LSn} lead to
value of
$-14.2~ kHz$.
Uncertainties arisen from one-logarithmic ones are
estimated as $8~ kHz$ for \eq{LSn} and $1~ kHz$ for \eq{LPn}. The
other
part of theoretical uncertainties is due to recoil corrections in order
$(Z\alpha)^6(m/M)m$. The uncertainty estimates is  $1~ kHz$ for \eq{LPn},
but the contribution to \eq{LSn} is zero \cite{PG}. That is why the uncertainty
for s-state difference
is smaller than in our works \cite{newkar,YF1,YF2}

According our evaluation above the highest precision result for the
$n=2$ Lamb splitting from measurements \cite{Sok}
differs from the original
result \cite{Sok} and from reevaluated one \cite{Pal}, both.

New
life time of $2p_{1/2}$ is now \cite{newkar,YF3}

\beq   \label{newtau}
\tau_{2p_{1/2}}=
1.5961887(15)\cdot 10^{-9}~ sec
,
\eeq

where the logarithmic correction \eq{newgam} leads to $-5.1\cdot 10^{-15}~
sec$ and our estimate of non-logarithmic $\alpha(Z\alpha)^2$
contributions is $1.5\cdot 10^{-15}~ sec$.

Using an
experimental value of product the life time and the $n=2$ Lamb
splitting measured in work \cite{Sok} a
new value of splitting is obtained \cite{newkar,YF3}
as

\beq  \label{newlamb}
L(2s_{1/2}-2p_{1/2})=1057.8576(21)~MHz,
\eeq

where the shift from result  of \cite{Sok}
without any radiative corrections
is
$-2.91~kHz$ and our estimate of
non-logarithmic contribution is $1~kHz$.
This result is in agreement with direct experimental values and with
theoretical one.

After summarizing items of \eq{way} according \eq{LSn}, \eq{LPn} and
\eq{newlamb} we obtain

\beq   \label {LS1}
\Delta E_L(1s_{1/2})
=
8172.934(22) ~MHz.
\eeq

This result has higher precision than both of direct experimental and
theoretical
ones.

A more detailed evaluation is presented in \cite{newkar} and
\cite{YF1,YF2,YF3}.

Let us discuss other Lamb shift values. Using Lundeen \& Pipkin result
\cite{Lun}
we can obtain

\beq   \label {LS1lp}
\Delta E_L(1s_{1/2})
=
8172.833(73) ~MHz,
\eeq

and  Hagley \& Pipkin one \cite{Hag} leads to

\beq   \label {LS1hp}
\Delta E_L(1s_{1/2})
=
8172.785(97) ~MHz.
\eeq

Using the difference \eq{LSn} we obtain from Garching measurement \cite{Wei}

\beq   \label {LS1G}
\Delta E_L(1s_{1/2})
=
8172.857(129) ~MHz,
\eeq

where 128 kHz leads from measurement
and theoretical uncertainty from \eq{LSn} is only 13 kHz.

We can re-evaluate results on $8s/d \to 2s$ \cite{Nez} and
$2s \to 1s$ \cite{And} two-photon transitions frequencies to obtain result
for 1s Hydrogen Lamb shift

\beq   \label {LS1GP}
\Delta E_L(1s_{1/2})
=
8172.786(118) ~MHz.
\eeq

All results for Hydrogenic atom are collected in the Table 1 and on the Fig. 1.
Some discussion of theoretical results is done lower with considering of
similar corrections to Muonium hyperfine splitting. We have incule into the
Table 1 and the Fig. 1 theoretical results with both proton radii and with and
without two-loop corrections. Result without
$\alpha^2(Z\alpha)^5m$ \cite{Pac2} is included only
because "with $\alpha^2(Z\alpha)^5m$ \cite{Pac2}"-result
is in huge disagreement with the Lamb
shift of the Helium-Ion
\cite{Wij}. It should be also mentioned that the large two-loop result is in
agreement
with measurements of hydrogen-like ions of the Phosphorus
\cite{Gas} and the of Sulfur \cite{Geo}.

The two-loop corrections in normalization of \cite{Pac2} are

\beq
H^{P}_{50}=-55(23), ~~~~from ~~ \cite{Geo},
\eeq

and

\beq
H^{S}_{50}=-62(36),~~ ~~ from ~~\cite{Gas},
\eeq

and theoretical values \cite{Pac2,kar} are

\beq
H_{50}(Z)=B_{50}
\left(
1 \pm (Z\alpha)
\right)
-\frac{8}{27
}
(Z\alpha)\ln^3{\frac{1}{(Z\alpha)^2}}
\left(
1 \pm 1
\right),
\eeq

or

\beq
H_{50}(Z=15) \approx H_{50}(Z=16) = - 24(4).
\eeq

The average value from the Chlorine and Argon is $H^{Cl \& Ar}_{50} = -85(74)$.

In this language the Helium disagreement is

\beq
H^{He}_{50}=0.2(4.6),~~ ~~ from ~~\cite{Wij}
\eeq

instead a theoretical value $H_{50}(Z=2) = - 27(3)$. The Lithium experimental
value is
$H^{Li}_{50} =  61(52)$.

Some theoretical discussion of two-loop corrections
is done some later.

\section{Results for the Deuterium}

We could also work for Deuterium s-state difference. The result is

\beq   \label {LSnd}
\Delta E_L(1s_{1/2}) -
8 \Delta E_L(1s_{1/2})
=
-187.229(8) ~MHz,
\eeq

or 8 kHz higher than for Hydrogen and  from Garching experiment \cite{Wei}
it leads to

\beq   \label {LS1Gd}
\Delta E_L(1s_{1/2})
=
8183.905(224) ~MHz,
\eeq

where uncertainty from \eq{LSnd} is only 13 kHz.

We can also obtain the Deuterim ground state Lamb shift using equations
\eq{LSn} and \eq{LSnd}
for s-state differences of Hydrogen and Deuterium, the (1s-2s)-isotope
shift measurement \cite{Schm}
and some result of the 1s Hydrogen Lamb shift. The value of \eq{LS1G}
leads to Garching-Garching
result

\beq   \label {LS1GGd}
\Delta E_L(1s_{1/2})
=
8183.967(132) ~MHz,
\eeq

which is in a good agreement with the direct Garching value \eq{LS1Gd}.

\section{Result for the Muonium Lamb shift}

The result for the Muonium is
\beq   \label {LSnm}
\Delta E_L(1s_{1/2}) -
8 \Delta E_L(1s_{1/2})
=
-187.348(8) ~MHz.
\eeq

\section{$\alpha^2Ry$ term and future measurements}

Real experimental values in so-called {\it Lamb shift} measurements are
rather Dirac correction to the Sch\"odinger energies. Dirac corrections
for Garching 1s Lamb shift experiment is 3928.707(12) MHz
where the uncertanty is experimental one.

The Paris-Garching result is

\beq
f(8d-2s)-\frac{5}{16}f(2s-1s) = 4187.518(18) MHz,
\eeq

where frequencies are double frequencies of two-photon transitions and
uncertainty items are 10 kHz (Paris) and 14 kHz (Garching).

Using these values one can obtain some fine structure constant values
with uncertainties like $(1.5 \div 2) \cdot 10^{-6}$. But at least
the Garching results are going to be improved and
to have uncertainties reduced
by factor 10
\cite{PCHan}. The Paris results are also going to be improved
\cite{PCCan}. That means that to calculate
 Dirac corrections should know the fine
structure constant with high precision.

\section{The fine structure constant}

There are a lot of way to determinate the fine structure constant. The main
experimental values are presented in the Table 2 and on the Figs. 2.
There are two
kinds of values there.
Some lead from electrical measurements and is due to electrical
standards. The other have no connections with them.
The main non-electrical values
are results from investigations of the  electron anomalous magnetic moment
\cite{amm,KinoL}

\beq
\alpha^{-1}_{AMM} = 137.035 9922(9)
,
\eeq

and one from the neutron  de Broglie's wave length measurements \cite{Kru}

\beq
\alpha^{-1}_n = 137.036 0105(54).
\eeq

The value from the photon recoil result for $h/M_{Cs}$ includes only
statistical error \cite{Cs}.

We are considered non-electrical results later. A detailed discussion
for electrical ones is presented in a review of
 Cohen and  Taylor \cite{Coh90}.

Equations to determinate  $\alpha^{-1}$ are follow:

\beq
\alpha^{-1}_{g-2}=\frac{a_e}{A_2+\frac{\alpha}{\pi}A_4+\dots}
,
\eeq

\beq
\alpha^{-2}_{Mu}=\frac{16}{3}(Ry\cdot c)(\mu_\mu/\mu_p)(\mu_p/\mu_B)(\Delta
\nu_{hfs}^{-1})(1+\dots)
,
\eeq

\beq
\alpha^{-2}_n=\frac{(m_n\cdot  c/h)}{2 \cdot Ry \cdot (m_n/m_e)}
,
\eeq

\beq
\alpha^{-2}_{Cs}=\frac{(M_{Cs}\cdot  c/h)}{2 \cdot  Ry \cdot (M_{Cs}/m_e)}
,
\eeq

\beq
\alpha^{-1}_{R_K,K_\Omega}=(2 \cdot R_K/\mu_0c)K_\Omega
,
\eeq

\beq
\alpha^{-2}_{\gamma_p',K_\Omega}=c\cdot (4\cdot Ry\cdot \gamma_p')^{-1}
(\mu_p'/\mu_B)(2e/h)
K_\Omega^{-1}
,
\eeq

\beq
\alpha^{-3}_{\gamma_p',R_K}=R_K(2\cdot \mu_0\cdot Ry\cdot \gamma_p')^{-1}
(\mu_p'/\mu_B)(2e/h)
,
\eeq

where $K_\Omega$ is a ratio of the SI Ohm and the BIPM Ohm, $R_K$ is the
Klitzing
constant which is expected to be equal to the Quantum Hall resistance,
$\gamma_p'$ is the giromagnetic ratio of the proton measured in the water
by low-field method, and all electrical values are measured in the BIPM units.
The QED corrections is presented as "$\dots$".

\section{The anomalous magnetic moment of electron}

The theoretical expession of anomalous magnetic moment of electron
has form

\beq
a_e = A_2
\frac{\alpha}{\pi}
+ A_4\frac{\alpha^2}{\pi^2}+ A_6\frac{\alpha^3}{\pi^3} + A_8
\frac{\alpha^4}{\pi^4} + \delta a,
\eeq

where electron-photon contributions are \cite{KinoL,Lap} (see also a review
\cite{Kinog-2})

\beq
A_2 (e) = 0.5
,
\eeq

\beq
A_4 (e) = - 0. 328 478 965 ...
,
\eeq

\beq
A_6 (e) = 1. 176 11 (42)
,\eeq

\beq
A_8 (e) = - 1. 424 (138)
{}.
\eeq

Heavy leptons contributions are

\beq
\delta a(\mu)    = 2.804  10^{-12},
\eeq

\beq
\delta a(\tau)    = 0.010  10^{-12}
{}.
\eeq

Non-QED corrections are also known

\beq
\delta a(had)  = 1.6(2)  10^{-12}
,\eeq

\beq
\delta a(weak) = 0.05    10^{-12}
{}.
\eeq

That leads to result

\beq
 a_e = 1 159 652 140 (27.1)(5.3)(4.1) \cdot 10^{-12},
\eeq

where we use $\alpha$ from {\it so-called} the Quantum Hall effect measurement
(see Table 2)

\beq
\alpha^{-1}_{R_K,K_\Omega} = 137.0359979(32).
\eeq

The items of uncertainties are 27.1 from $\alpha$, 5.3 from $A_6$ and 4.1 from
$A_8$.

The results for the fine structure constant is

\beq
\alpha^{-1}_{g-2} = 137. 035 992 22 (63)(48)(51)
\eeq

where two first uncertainties are due  $A_6$ and $A_8$, the last
arrisen from the measurements \cite{amm}

\beq
 a_{e^-} =
1159 652 187.9 (4.3)  10^{-12}
\eeq

and

\beq
 a_{e^+} =
1159 652 188.4 (4.3)  10^{-12}.
\eeq

\section{The Muonium hyperfine splitting}

The main contribution to the Muonium ground state hyperfine splitting can
be writen as

\beq  \label{EF}
  \nu_F=\frac{16}{3}{(Z\alpha)}^2 c R_{\infty} \frac{m}{M}
  \left [ \frac{m_R}{m} \right ]^{3}(1 + a_{\mu} ).
\eeq

The quantum electrodynamics part of the HFS interval is
\cite{Sap,STY,BYG,EKSe,EKSm}

\beq  \label{defth}
  \Delta \nu ({QED}) =
\nu_F
\Big(
1 +
\frac{3}{2} (Z \alpha )^2
+\frac{17}{8} (Z \alpha )^4 + \dots
\Big)
\eeq

\[
+
\nu_F  \Big(
a_e + \alpha \Big\{ (Z \alpha )( \ln 2 - \frac{5}{2} )
\]

\[
 - \frac{8 (Z \alpha )^2}{3 \pi} \ln (Z\alpha)
\left[ \ln (Z\alpha) - \ln 4 + \frac{281}{480} \right]
\]

\[
+ \frac{(Z\alpha )^2}{\pi} (15.38 \pm 0.29 )
\Big)\Big\}
+\frac{\alpha^2(Z\alpha)}{\pi}D
\Big)
\]

\[
+
\frac{\nu_F}{1 + a_{\mu} }
\Big( - \frac{3Z\alpha}{\pi}
\frac{m M} {M^2 -m^2} \ln {\frac{M}{m}}
\]

\[
+
\frac{\gamma^2}{m M}
\left [ 2 \ln {\frac{1}{2Z\alpha}}
- 6 \ln 2 + \frac{65}{18} \right ] \Big)
\]

\[
+
\frac{\nu_F}{1 + a_{\mu} }
\frac{\alpha (Z\alpha)}{\pi^2} \frac{m}{M}
\Big( - 2 \ln^2
{\frac{M}{m}}
 + \frac{13}{12} \ln {\frac{M}{m}}
\]

\[
+ \frac{21}{2} \zeta (3) + \frac{\pi^2}{6} + \frac{35}{9}\Big)
\]

\[
+
\nu_F
\frac{\alpha^2 (Z\alpha)}{\pi^3}\frac{m}{M}
\Big(
- \frac{4}{3} \ln^3 {\frac{M}{m}}
+ \frac{4}{3} \ln^2 {\frac{M}{m}}
 + \dots
 \Big),
\]

where D is two-loop coefficient. Two loop corrections and higher order
conttributions are considered in next sections.

Non-QED corrections are known \cite{Kari}

\beq    \label{strong}
  \Delta \nu ({strong})
=
\nu_F
\frac{\alpha (Z\alpha)}{\pi^2}
\frac{m}{M}
(2.15 \pm 0.14 ),
\eeq

and \cite{Beg,Bodweak}

\beq \label{weak}
  \Delta \nu ({weak}) \simeq  0.065 kHz.
\eeq

\section{Two-loop corrections}

Two-loop corrections of absolute order $\alpha^2 (Z\alpha)^5$ are indaced by
six gauge-independant sets of diagrams (see e.g.\cite{EKS1,EGO}) which
are presented in Fig. 3.

The contribution to the HFS interval in the Muonium ground state are:

\beq \label{1a}
  \Delta \nu^{a} =
\nu_F
\frac{\alpha^2 (Z\alpha )}{\pi}
\frac{36}{35},~~~~~~~\cite{EKS1}
\eeq

\beq   \label{1b}
  \Delta \nu^{b} =
\nu_F
\frac{\alpha^2 (Z\alpha )}{\pi}
\left (
\frac{224}{15}
\ln 2 - \frac{38}{15}\pi -\frac{118}{225} \right )
,~~~~~~~\cite{EKS1}
\eeq

\beq \label{1c}
  \Delta \nu^{c}=
\nu_F
\frac{\alpha^2 (Z\alpha )}{\pi}
\Big (
 -\frac{4}{3} {\cal L}^2  -\frac{20\sqrt{5}}{9} {\cal L} - \frac{64}
 {45} \ln 2
\eeq

\[
+ \frac{\pi^2}{9} + \frac{1043}{675} + \frac{3}{8}
 \Big ),~~~~~~~\cite{EKS1}
\]

where

\[
{\cal L} = \ln{\frac{1 + \sqrt{5}}{2}},
\]

\beq \label{fig1d}
\Delta \nu^{d} =
\nu_F
\frac{\alpha^2 (Z\alpha )}{\pi}
\left(
-0.310~742 \dots
\right).~~~~~~~\cite{EKS4}
\eeq

These results have been independently checked in the work \cite{Kinohfs}.

The  light-by-light scattering diagram leaded to slight disagreement
between  \cite{EKS5}  and \cite{Kinohfs}, which was due the misprint
on an intermediate state of work \cite{EKS5} (see  \cite{EKS5E}). The final
result is

\beq \label{1e}
  \Delta \nu^{(e)} =
\nu_F
\frac{\alpha^2 (Z\alpha )}{\pi}
\left(
-0.47251
\dots
\right).~~~~~~~\cite{EKS5,EKS5E},~~\cite{Kinohfs}
\eeq

The result of the sixth set calculation is

\beq \label{1f}
  \Delta \nu^{f} =
\nu_F
\frac{\alpha^2 (Z\alpha )}{\pi}
\left(
-0.63(4)
\right).~~~~~~~\cite{Kinohfs}
\eeq

This figure was obtained in Feynman gauge.  An other way is used in works
\cite{EKS61,EKS62}, where the first nine diagram of this nineteen-diagram set
(see Fig.4)
were evaluated. Evaluation is done in Fried-Yennie gauge. The result for sum of
three first
diagram of \cite{Kinohfs} is in agreement with earlier result of \cite{EKS61}.
The calculations in Fried-Yennie gauge are going to be completed \cite{ES63}.

\section{Two-loop corrections. Comparision: HFS and Lamb shift}

The contributions to HFS interval can be writen as

\beq                                  \label{Fhfs}
\Delta \nu_{hfs} = \nu_F
\frac{
\alpha^2
(Z\alpha)
}{\pi n^3}
\left(
\frac{8}{\pi^2}
\int_0^\infty\frac{d
\vert {\bf k} \vert }
{{\bf k}^2}
F_{hfs}({\bf k})
\right),
\eeq

where $F_{hfs}({\bf k})$ is some substructed form-factor, normatized as 1 in
the
sceleton diagram.
The substraction need to exlude the lower order corrections.

The contributions to Lamb shift can be writen as

\beq                                  \label{Fls}
\Delta \nu_{Ls} =
\frac{
\alpha^2
(Z\alpha)^5m
}{\pi^2 n^3}
\left(
-\frac{16}{\pi}
\int_0^\infty\frac{d
\vert {\bf k} \vert }
{{\bf k}^4}
F_{Ls}({\bf k})
\right),
\eeq

where $F_{hfs}({\bf k})$ is some substructed form-factor, normatized also as 1
in the
sceleton diagram.
The substraction need also to exlude the lower order corrections.

To compare contribution let us consider integrals

\beq                                  \label{I}
I =
\frac{8}{\pi}
\int_0^\infty\frac{d
\vert {\bf k} \vert }
{{\bf k}^{2r}}
F_{Ls}({\bf k})
,
\eeq

where $r_{hfs}=1$ and $r_{Ls}=2$. In Table 3 we include the numerial results
for all $I$, and asymptotic behaviours of integrand  $F({\bf k})/{\bf k}^{2r}$.
For non-leading degree of $k$ no logarithmic factors
are presented.

Taking into account that the integration of logarithm should lead to higher
result,
we can expect that Lamb shift contribution of a, b, c, d, e should be smaller,
but f-corrections
should be larger. Realy such estimate could be not correct if constant is quite
large, as it is in
two-loop vacuum polarization (Fig. 3c).
But in this case we have additional reason to expect lower contribution to
the Lamb shift. The $k^4$-term in low-energy expanssion of polarization is
product
of $k^2$ term and
$k^2/(2m)^2$ and some constant, which is smaller than 1. So after substracting
the low-energy part of integral
should be smaller.

Considering the Pachucki result \cite{Pac2}, it should be mentioned that
integrand has correct
asymptotic behaviours
for both low and high momenta,
and integrating the low-energy asymptotics from 0 to the electron mass and the
high-energy one from twice electron mass to the infinity we can obtain

\beq
I^f_{Ls}=(7 \div 14) + 3,
\eeq

where to estimate uncertainty we use at low-energy

\beq
\Big( \ln{1/k} - 1 \pm 1 \Big)
\eeq

instead $\ln{1/k}$.

We expect that the asymptotics is rather
some over-estimate of the real integrand and
so a naturel magnitude of this integral is $10 \pm 5$.

In work \cite{Pac2} the Fried-Yennie gauge was used for the first three
diagram of Fig 4, and Feynman gauge for the other graphs.
The result of three first
diagram of \cite{Pac2} is in agreement with earlier result of \cite{EKS61},
where the first nine diagram of this nineteen-diagram set (see Fig.4)
were evaluated. Evaluation is done in Fried-Yennie gauge.
The calculations are going to be completed \cite{EKS62,ES63}.

\section{Higher order logarithmic corrections}

The leading logarithmic radiative-recoil corrections are known

\beq
\delta \nu^{(1)}(rad-rec)=\nu_F
\frac{\alpha^2(Z\alpha)}{\pi^3}
\frac{m}{M}
\left(
-\frac{4}{3}
\ln^3{\frac{M}{m}}+
\frac{4}{3}
\ln^2{\frac{M}{m}}\right)
\eeq

The cubic term was obtained \cite{ES}, and square of recoil logarism was
calculated in \cite{EKSlog}. Some additional discussion on light-by-light
scattering contribution was done in in \cite{KSElog5}, and on diagrams
without closed electron loop in in \cite{Mugau}. The contributions of these
recoil terms are -0.04 and +0.01 kHz.

More valuable corrections include low-energy logarithm. The well-known
leading corrections in order $\alpha(Z\alpha)^2$  can be obtained without
using any expanssion of the mass ratio $m/M$. The result is \cite{kar}

\beq                  \label{mai}
\delta \nu({leading})=
\nu_F
\frac{\alpha
}
{\pi}
(Z\alpha)^2
 \left(
-
\frac{8
}
{3}
 \left(
\frac{m_R}{m}
\right)^2
\ln^2{Z\alpha}
\right)
{}.
\eeq

Hence, a new radiative-recoil correction is \cite{kar}

\beq                  \label{rrclog}
\delta \nu^{(2)}({rad-rec})=
\nu_F
\frac{\alpha
}
{\pi}
(Z\alpha)^2
\frac{m}{M}
 \left(
+
\frac{16
}
{3}
\ln^2{Z\alpha}
\right)
,
\eeq

or 0.34 kHz. The contribution of the electron anomalous magnetic moment
is \cite{kar}

\beq                  \label{radlog}
\delta \nu({rad})=\nu_F
\left(
\frac{\alpha}{\pi}
\right)^2
(Z\alpha)^2
\left(
-\frac{4}{3}
\ln^2{Z\alpha}
\right)
,
\eeq

or -.04 kHz.

Taking into accout the recoil correction to the Lamb shift instead
the pure radiative contribution we can also obtain a recoil contribution
\cite{kar}

\beq                  \label{reclog1}
\delta \nu^{(1)}({rec})= \nu_F
\frac{
(Z\alpha)^3}
{
\pi}
\frac{m}{M}
\left(
-
\frac{2}{3}
\ln^2{Z\alpha}
\right)
,
\eeq

or -.04 kHz. However, there are an other logarithmic contribution. This can be
evaluated from the leading recoil contribution but with the Dirac wave
functions
\cite{25EGAS,CPEM94}

\beq                \label{reclog2}
\delta \nu^{(2)}({rec})=
\nu_F
\frac{(Z\alpha)^3}{\pi}
\frac{m}{M}
\left(
-3\ln{\frac{M}{m}}
\ln{\frac{1}{Z\alpha}}
\right),
\eeq

the magnitude of corrections is -.21 kHz

The leading higher order correction to the Lamb shift  is \cite{kar}

\beq
\delta E^{cub}_L(ns)=-\frac{8}{27
}\frac{\alpha^2 (Z\alpha)^6m}{
\pi^2
n^3}\ln^3{\frac{1}{(Z\alpha)^2}},
\eeq

or -3.6 kHz for 2s, -29 kHz for 1s Lamb shift. This corrections is due to the
same
set of diagram
(Fig. 3f)
as conntribution \cite{Pac} and it is about 9\% of it.

The part of logarithm square corrections arises from set of Fig 3c.

\beq
\delta E^{sq,f}_L(ns)=\frac{4}{45
}\frac{\alpha^2 (Z\alpha)^6m}{
\pi^2
n^3}\ln^2{\frac{1}{(Z\alpha)^2}},
\eeq

or .88 kHz for 1s Lamb shift. This corrections is about 4\% of
$\alpha^2(Z\alpha)^5m$
contribution
of diagram Fig. 3c.

\section{The Muonium and the fine structure constant}

To determinate the fine structure constant we sould re-wrire the  theoretical
equation using only values which can be measured

\beq           \label{alp}
\Delta \nu (th) =
\frac{16}{3}\alpha^2 c Ry
\left(
\frac{\mu_\mu}{\mu_p}
\frac{\mu_p}{\mu_B}
\right)
\left [ \frac{m_R}{m} \right ]^{3}
\left\{ 1 + \dots
\right\},
\eeq

where the main problem is due to the muon magnetic moment. All known result
are \cite{Kle,Mar} and references there
in good agreement but they have not quite high precision (Fog 5). The new
result
is going to have uncertainty reduced by factor like $3 \div 5$ and relative
uncertainty for $\alpha$ will be $(3 \div 5)\cdot 10^{-8}$.

The final theoretical result for the HFS interval is

\beq
\Delta \nu (th) = 4463303.6 (13)(2) kHz,
\eeq

where we use $\alpha$ from the anomalous magnetic moment of electron, and
uncertainty
items due to
the muon magnetic moment (average value of \cite{Kle,Mar} and theoretical
calculations
(numerical error
of integration in \cite{Sapi} and an estimate of higher-order uncalculated
contributions).

The highest precision experimental result is

\beq
\Delta \nu (exp) = 4463302.88 (16) kHz.
\eeq

\section{The neutron  de Broglie's wave length}

The neutron result of $\alpha$ is

\beq
\alpha^{-1}_n = 137.035 993(27).
\eeq

It obtained from experimental results for the Rydberg constant, the
neutron-electron masses ratio, the neutron de Broglie's wave length $(h/m_n
v)$ \cite{Kru}, which was measured in unity of some known crystal lattice
spacing, and
the neutron velocity $v$ \cite{Kru}:

\beq                                \label{alm}
\alpha_n = \sqrt{\frac{c Ry (m_n/m_e)}{(m_n v/2 h) v}}.
\eeq

This result is in disagreement with the one from the electron  anomalous
magnetic
moment

\beq                   \label{amm}
\alpha^{-1}_{AMM} = 137.035 9922(9).
\eeq

The use of a crystal leads to a direct connection between the Sillicon spacing
measured
and the fine structure measurements. We can use \eq{alm} and \eq{amm} and
obtain
indirect result for
the Sillicon spacing

\beq
d_{220}(indirect) =192 015.617(10)~fm
\eeq

instead PTB direct result \cite{PTB}

\beq
d_{220}(direct) =192 015.568(12)~fm.
\eeq

Recenlty a new result for the spacing have been obtained \cite{Bas}. It is
equal to

\beq
d_{220} =192 015.569(6)~fm.
\eeq

So we can expect that after some re-evaluating,
the neutron result for the fine structure constant should have the same value
but twice smaller
uncertainty (its relative value will be $\approx 2 \cdot 10^{-8}$). The one for
the anomalous moment is three times smaller,
but disagreement is $\approx 1.3 \cdot 10^{-7}$.

All values of the Sillicon lattice spasing \cite{PTB,Bas,Des} are presented on
Fig. 6.

We also expect that some connection between Avogadro constant
measurements and the fine structure constant
could appeare. Generally, the results for $\alpha$ are obtained
with higher precision, but there is a factor 6
between uncertainty values. That is because the fine structure
constant is connected with square root of the spacing and
the Avogadro constant is proportional to inverse cube of it.

The other problem is due the neutron-electron mass ratio.
The neutron mass is result of measurements of the proton mass,
Deuteron mass and the Deuteron binding energy.
For the last value the highest precision result \cite{Gre} has some not very
large
disagreement with others \cite{Bind} (see Fig. 7). The shift of $\alpha$ could
be $10^{-8}.$

\appendix{\bf Acknowledgement}

A part of this work was done during my stay at the Physikalisch -
Technische Bundesanstalt (Braunschweig) at summer, 1993 and I would like
to thank the PTB and for support and hospitality.

I am very grateful
H.  Bachmair,   P. Becker,
E. Braun,  B. Cagnac,  R.
Damburg, G. W. F. Drake, M.  I. Eides, R., N. Faustov,
 A. Huber,
 A. I. Ivanov,
 V. G. Ivanov, K. Jungmann, I. B.  Khriplovich,  W. K\"onig, V. Koze, B.
Kramer,
E. Kr\"uger, L. N.  Labzobsky, D. Leibfried, G. P. Lepage,  S. R. Lundeen,  N.
L.
Manakov, V. G. Pal'chikov, H.  Rinneberg, V. M. Shabaev, V.  A.  Shelyuto,
V. S. Tuninsky,  W. Weirauch, G.-D.
Willenberg, M. Weitz, W. W\"oger and  V. P. Yakovlev
for
stimulating and fruitful discussions.

I would like especially to knowledge the Max-Plank-Institut f\"ur
Quantenoptik for hospitalily at the automn, 1994 during a final
part of this work and T. W. H\"ansch
and  K. Pachucki for useful discussion and information on their yet
unpublished and uncompleted results.

I would like  also to thank G. zu Putlitz and K. Jungmann for invitation to
give
a seminar at and at the Physikalisches Institut
(the Universit\"at Heidelberg) and for hospitalily.

The work described above was possible in part after support of
the International Science Foundation (Grant \#R3G000).

\newpage

\begin{center}
{\bf
Table 1. The Lamb shift of  the Hydrogen ground
state
}

\vspace*{5mm}
\begin{tabular}{||c|c||}
\hline
\hline
$Ref.$&Result\\
\hline
Lundeen \& Pipkin&
8172.833(73) MHz\\
\hline
Sokolov \& Yakovlev (corrected)&
8172.934(22) MHz\\
\hline
Hagley \& Pipkin&
8172.785(97) MHz\\
\hline
Garching&
8172.857(129) MHz\\
\hline
Garching \& Paris&
8172.786(118) MHz\\
\hline
$r_p=.862 fm, \alpha^2(Z\alpha)^5m$&
8172.762(40) MHz\\
\hline
$r_p=.862 fm$&
8173.053(40) MHz\\
\hline
$r_p=.805 fm, \alpha^2(Z\alpha)^5m$&
8172.613(40) MHz\\
\hline
$r_p=.805 fm$&
8172.904(40) MHz\\
\hline
\hline
\end{tabular}
\end{center}
\vspace*{5mm}

\begin{center}
{\bf
Table 2. The (inverse) fine structure constant $\alpha^{-1}$
}

\vspace*{5mm}
\begin{tabular}{||c|c|c||}
\hline
\hline
$NN$.&Method&Result\\
\hline
1.&$(g_e-2)$&
137.035 9922(9)\\
\hline
2.&$R_K,K_\Omega,NIST$&
137.035 9979(33)\\
\hline
3.&$\gamma_p',R_K,NIST$&
137.035 9840(51)\\
\hline
4.&$h/m_n$&
137.0360102(54)\\
\hline
5.&$h/M_{Cs}$&
137.0360876(71)\\
\hline
6.&$R_K,K_\Omega,NPL$&
137.0360084(74)\\
\hline
7.&$R_K,K_\Omega,CSIRO/NML$&
137.0360093(90)\\
\hline
8.&$\gamma_p',R_K,VNIIM$&
137.035 949(16)\\
\hline
9.&$Mu hfs$&
137.036003(20)\\
\hline
10.&$Cohen,Taylor,1992$&
137.035 9928(9)\\
\hline
11.&$CODATA,1986$&
137.035 9986(62)\\
\hline
12.&$CCE,1990$&
137.035 997(27)\\
\hline
\hline
\end{tabular}
\end{center}
\vspace*{5mm}

\newpage

\begin{center}
{\bf Table 3a.\\
The two-loop values of HFS \cite{EKS1}, \cite{EKS4}, \cite{EKS5,EKS5E},
\cite{Kinohfs}
}

\vspace*{5mm}
\begin{tabular}{||c|c|c|c|c|c|c|c||}
\hline
\hline
$NN.$&a&b&c&d&e&f&tot\\
\hline
$I$&3.24&5.87&-2.10&-.98&
-1.48&-2.0&2.56
\\
\hline
$k\ll m$&$k^2$&1&1&1&$k^2$&$\log{1/k}$&-
\\
\hline
$k\gg
m$&$\log^2{k}/k^2$&$\ln{k}/k^2$&$\log{k}/k^2$&$\log{k}/k^2$&$\log{k}/k^2$&$1/k^2$&-
\\
\hline
\hline
\end{tabular}
\end{center}
\vspace*{5mm}

\begin{center}
{\bf Table 3b.\\
The two-loop values of Lamb shift, \cite{EGO}, \cite{EG3}, \cite{EG4},
\cite{Pac1}, \cite{EGP}, \cite{Pac2}
}

\vspace*{5mm}
\begin{tabular}{||c|c|c|c|c|c|c|c||}
\hline
\hline
$NN.$&a&b&c&d&e&f&tot\\
\hline
$I$&.096&-.80&-.96&.11&
.19&12.0&10.6
\\
\hline
$k\ll m$&1&1&$\log{1/k}$&1&1&$\log^2{1/k}$&-
\\
\hline
$k\gg m$&$1/k^4$&$1/k^2$&$1/k^4$&$1/k^2$&$1/k^4$&$1/k^2$&-
\\
\hline
\hline
\end{tabular}
\end{center}
\vspace*{5mm}

\end{document}